\documentclass[final,5p,times,twocolumn]{elsarticle}

\usepackage{graphicx}
\usepackage{amssymb}
\usepackage{amsmath}
\usepackage{hyperref}
\journal{Physics Letters B}

\begin{document}

\begin{frontmatter}



\title{Momentum Spectra for Dynamically Assisted Schwinger Pair Production}

\author[kfuni,mcl]{M. Orthaber}%\corref{cor1}}
%\ead{markus.orthaber@mcl.at}
\author[kfuni]{F. Hebenstreit}
%\ead{florian.hebenstreit@uni-graz.at}
\author[kfuni]{R. Alkofer}
%\ead{reinhard.alkofer@uni-graz.at}

\address[kfuni]{Institut f\"ur Physik, Karl-Franzens-Universit\"at Graz, A-8010 Graz,
Austria}
\address[mcl]{Materials Center Leoben Forschung GmbH., A-8700 Leoben,
Austria}
\begin{abstract}
Recently the dynamically assisted Schwinger mechanism, i.e., 
electron-positron pair production from vacuum by a combination of laser pulses 
with different time scales has been proposed. 
The corresponding results, which suggest that the rate of produced pairs is 
significantly enhanced by dynamical effects, are verified. 
Employing the framework of quantum kinetic theory
intrinsically enables us to additionally provide momentum space 
information on the generated positron spectrum.
\end{abstract}


\end{frontmatter}

\section{Introduction}

The description of a charged particle within Quantum Electrodynamics (QED) necessitates to take the corresponding antiparticle into account. Historically, the first example has been the prediction of the positron, the antiparticle of the electron. One may exploit the similarity of quantum field theory to many-body physics by introducing the Dirac sea picture: The vacuum is modeled by a sea of fermions with all negative-energy levels filled. The antiparticle is then represented by a hole in this sea. This makes plain that an antiparticle-particle pair can be created as soon as enough energy is available to bridge the mass gap. 

It became evident soon that one can create an antiparticle-particle pair either perturbatively from photons in high-energy reactions or non-perturbatively in the presence of very strong electric fields. The latter type has been called Schwinger mechanism \cite{Schwinger:1951nm} despite the fact that it has been discussed already many years before 
\cite{Sauter:1931zz,Heisenberg:1935qt}. This effect has attracted a lot of
interest recently
\cite{Alkofer:2001ik,Ringwald:2001ib,Blaschke:2005hs,Schutzhold:2008pz,Dunne:2009gi,Hebenstreit:2009km,Hebenstreit:2008ae,Ruf:2009zz,DiPiazza:2009py,Bulanov:2010ei} since laser facilities such as the Extreme Light Infrastructure (ELI) or the European XFEL might possibly reach the required field strengths. 

Being aware of the possibilities of perturbative versus non-perturbative production mechanisms, the question arises how to switch in a quasi-continuous way from one to the other. In this respect it turns out that time-dependent strong electric fields do this when varying the relevant frequency. Consequently, allowing for different and suitably chosen frequencies leads to a kind of \textit{dynamically assisted Schwinger mechanism} \cite{Schutzhold:2008pz}. 

The aim of the investigation reported in this letter is twofold: First, the mechanism as such is verified employing a quantum kinetic equation instead of the (semi-classical) world-line formalism. Second, and in view of the experiments even more important, momentum spectra of the produced pairs are predicted. Whilst the considered case is certainly not realistic in the sense of representing experimentally relevant superpositions of electric fields, it already demonstrates very nicely what kind of visible effects in the momentum spectra are to be expected in appropriately tuned upcoming experiments.

This letter is structured as follows: In Sect.~\ref{sec2} we set the theoretical basis for our investigations and introduce a simple model for dynamically assisted Schwinger mechanism\footnote{For a detailed and self-contained description, the interested reader may refer to \cite{Orthaber:2010}.} which is based on the proposal \cite{Schutzhold:2008pz}. In Sect.~\ref{sec3} we discuss our numerical results on the momentum spectrum as well as on the particle density. We show that combining the mechanisms of perturbative and non-perturbative pair creation results in strong non-linear behavior which in turn leads to an enhancement in the particle yield.

\section{Pair Production in Multiple Time Scale Fields}
\label{sec2}

\subsection{Homogeneous Background Field Approximation}

With the advent of high-intensity laser facilities such as ELI or the European
XFEL, we hope that it will become possible to create field strengths of the
order of $\mathcal{E}_{c} = {m^2}/e \sim 10^{16}V/cm$ in the focus of
colliding laser pulses. A laser should be described as an ensemble of photons in
principle, showing both temporal and spatial variation. However, due to the fact
that a complete description of this system is not yet feasible, we will make several reasonable simplifications in our calculations:
\begin{itemize}
  \item{We consider very high field strengths. Consequently, it is a reasonable assumption to treat the laser as classical background field. We thus neglect processes which could lead to a possible depletion of the laser due to electromagnetic cascades as indicated recently \cite{Bell:2008zzb,Fedotov:2010ja,Bulanov:2010gb}.}
  \item{The Schwinger mechanism is driven by the electric field. Having in mind
  a situation of colliding laser pulses with canceling magnetic fields, we will
  ignore the magnetic field completely; for studies including magnetic field effects see e.g. \cite{Kim:2003qp, Tanji:2008ku}.\\}
  \item{The electric field is space- and time-dependent in general. Nevertheless, since the achievable spatial focusing scale is orders of magnitude larger than the Compton wavelength, we will ignore any spatial dependence. For studies on the spatial dependence of the Schwinger mechanism see e.g. \cite{Gies:2005bz,Dunne:2006ur,Kleinert:2008sj,Hebenstreit:2010vz}.}
\end{itemize}
Consequently, we assume a spatially homogeneous electric field which is represented by a vector potential in temporal gauge $A_{\mu}(t) =(0,\mathbf{A}(t))$. The spatial part is chosen to point into the $\mathbf{e}_3$-direction: $ \mathbf{A}(t) = (0,0,A(t))$, such that the magnetic field vanishes, $ \mathbf{B}(t) = \nabla \times \mathbf{A}(t) = 0 $, and the electric field reads:
\begin{align}
  \mathbf{E}(t) = - \frac{d\mathbf{A}(t)}{dt} = (0,0,E(t)).
\end{align}

\subsection{Quantum Kinetic Formalism}

We employ the quantum kinetic formalism in order to obtain not only production rates but also to gain information on the momentum spectrum of created particles. This additional information is highly relevant, particularly with regard to a possible future detection of the Schwinger mechanism. The spectral information is encoded in the distribution function $f(\mathbf{q},t)$: It has to be stressed that there is no clear interpretation of $f(\mathbf{q},t)$ as long as the electric field is present. At finite times it cannot be considered as distribution function of real particles but only as a mixture of real and virtual excitations. Consequently, $f(\mathbf{q},t)$ might be interpreted as the momentum distribution for real particles only at \textit{asymptotic times} $t\to\pm\infty$, when the external field is switched off. This interpretation is also supported by S-matrix theory which clearly states that the identification of excitations of quantum fields with its corresponding particles is \emph{only} possible at asymptotic but not at intermediate times. Alternative interpretations, which do not account for this peculiarity of $f(\mathbf{q},t)$, lead to somewhat astonishing results \cite{Blaschke:2005hs,Gregori:2010uf}.

The equation of motion for $f(\mathbf{q},t)$ can be derived from canonical quantization, quantizing the spinor field fully but considering the electromagnetic field as given background \cite{Kluger:1998bm,Smolyansky:1997fc,Schmidt:1998vi}. Alternatively, a derivation based on a Hartree approximation of the relativistic Wigner function for QED is possible \cite{Hebenstreit:2010vz,BialynickiBirula:1991tx}. Due to the fact that we are mostly interested in the sub-critical field strength regime $\mathcal{E}\ll \mathcal{E}_c$ where the expected densities are rather low, we neglect collisions of created particles \cite{Tanji:2008ku,Vinnik:2001qd}. Additionally, the self-consistent field current due to created particles, which leads to an internal electric field and the decrease of the background, should be taken into account in principle. Detailed analysis, however, showed that these contribution can be safely neglected in the sub-critical field strength regime $\mathcal{E}\ll \mathcal{E}_{c}$ as well \cite{Tanji:2008ku,Bloch:1999eu,Roberts:2000aa}. With these simplifications, the quantum kinetic equation for $f(\mathbf{q},t)$ reads:
\begin{equation}
  \label{qk_int}   
  \dot{f}(\mathbf{q},t)=W(\mathbf{q},t)\int\limits_{-\infty}^{t}{dt'W(\mathbf{q},t') 
  \left[1-f(\mathbf{q},t')\right]\cos\left[2 \Theta(\mathbf{q};t,t')  \right]} \ ,
\end{equation}
where $f(\mathbf{q},t)$ accounts for both spin directions due to the absence of magnetic fields. Here,

\begin{equation}
  W(\mathbf{q},t)=\frac{eE(t)\epsilon_\perp}{\omega^2(\mathbf{q},t)} \quad \mathrm{and} \quad \Theta(\mathbf{q};t_1,t_2)=\int\limits_{t_1}^{t_2} dt' \omega(\mathbf{q},t') \ \nonumber  ,
\end{equation}
with $\mathbf{q}=(\mathbf{q}_\perp,q_3)$ being the canonical momentum; $e$ denotes the electron charge; $\epsilon^2_{\perp}=m^2+\mathbf{q}^2_{\perp}$ is the transverse energy squared, whereas $\omega^2(\mathbf{q},t)= \epsilon^2_{\perp}+p_3(t)^2$ is the total energy squared\footnote{The well known connection between canonical and kinetic momentum is $\mathbf{p}(t) = \mathbf{q} - e\mathbf{A}(t)$.}. Alternatively, this equation may be expressed as linear, first order, ordinary differential equation (ODE) system \cite{Bloch:1999eu}:
\begin{eqnarray}
  \dot{f}(\mathbf{q},t)&=&W(\mathbf{q},t)v(\mathbf{q},t) \nonumber\\     
  \dot{v}(\mathbf{q},t)&=&W(\mathbf{q},t)[1-f(\mathbf{q},t)]-2\omega(\mathbf{q},t)u(\mathbf{q},t)
  \label{qk_diff} \\
  \dot{u}(\mathbf{q},t)&=&2\omega(\mathbf{q},t)v(\mathbf{q},t) \nonumber
\end{eqnarray}
with initial conditions $f(\mathbf{q},t_i)=v(\mathbf{q},t_i)=u(\mathbf{q},t_i)=0$ at $t_i\to-\infty$. Both the integro-differential equation (\ref{qk_int}) and the ODE system (\ref{qk_diff}) are valid for any time-dependent electric field. 

In addition to the spectral information, the second important observable is given by the asymptotic particle number density $n[e^+e^-]$, which is calculated directly from $f(\mathbf{q},t)$:

\begin{equation}
  n[e^+e^-]=\int{\frac{d^3q}{(2\pi)^3}f(\mathbf{q},\infty)} \ . 
  \label{n+-}
\end{equation}

\subsection{Keldysh Parameter}

\begin{figure}[t!]
\begin{center}
  \begin{minipage}[b]{4.38 cm}
     \includegraphics[width = 125pt,height=61pt]{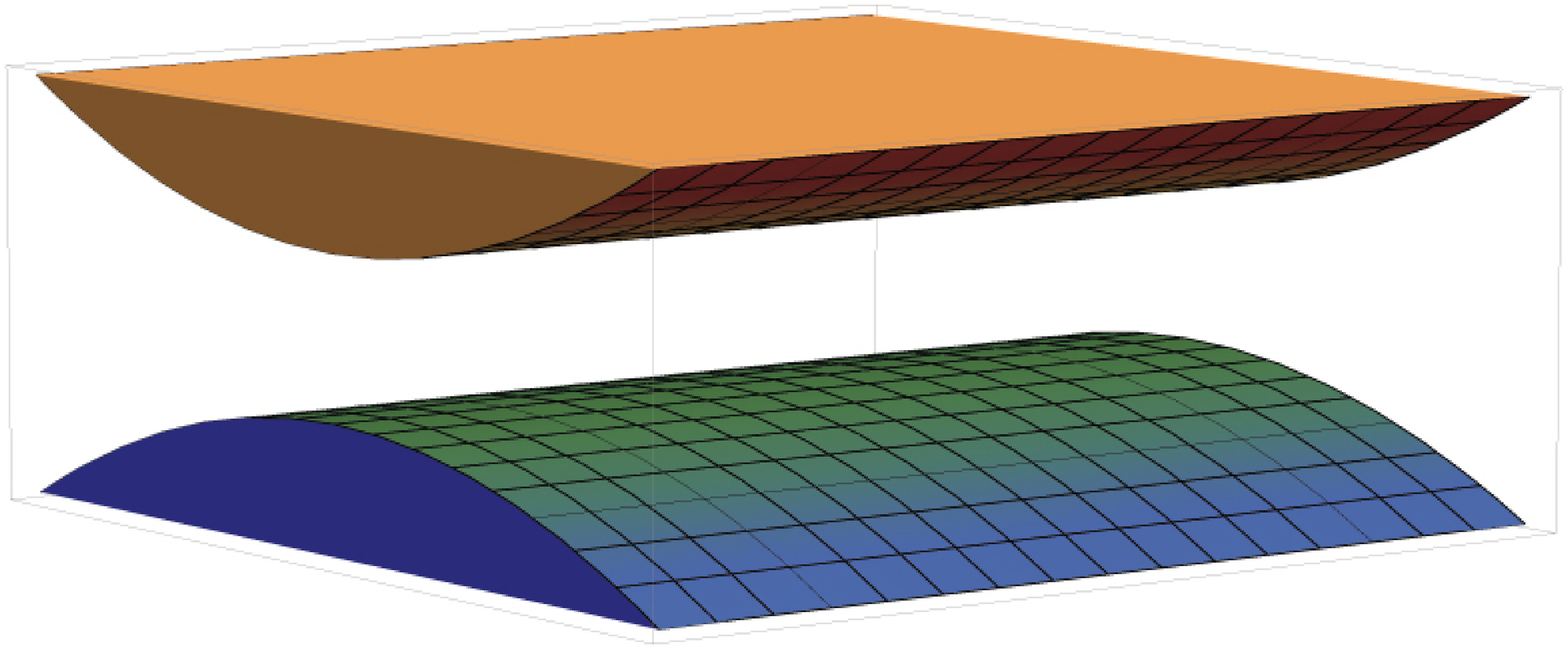}
   \end{minipage}
 \begin{minipage}[b]{4.38 cm}
     \includegraphics[width = 125pt,height=61pt]{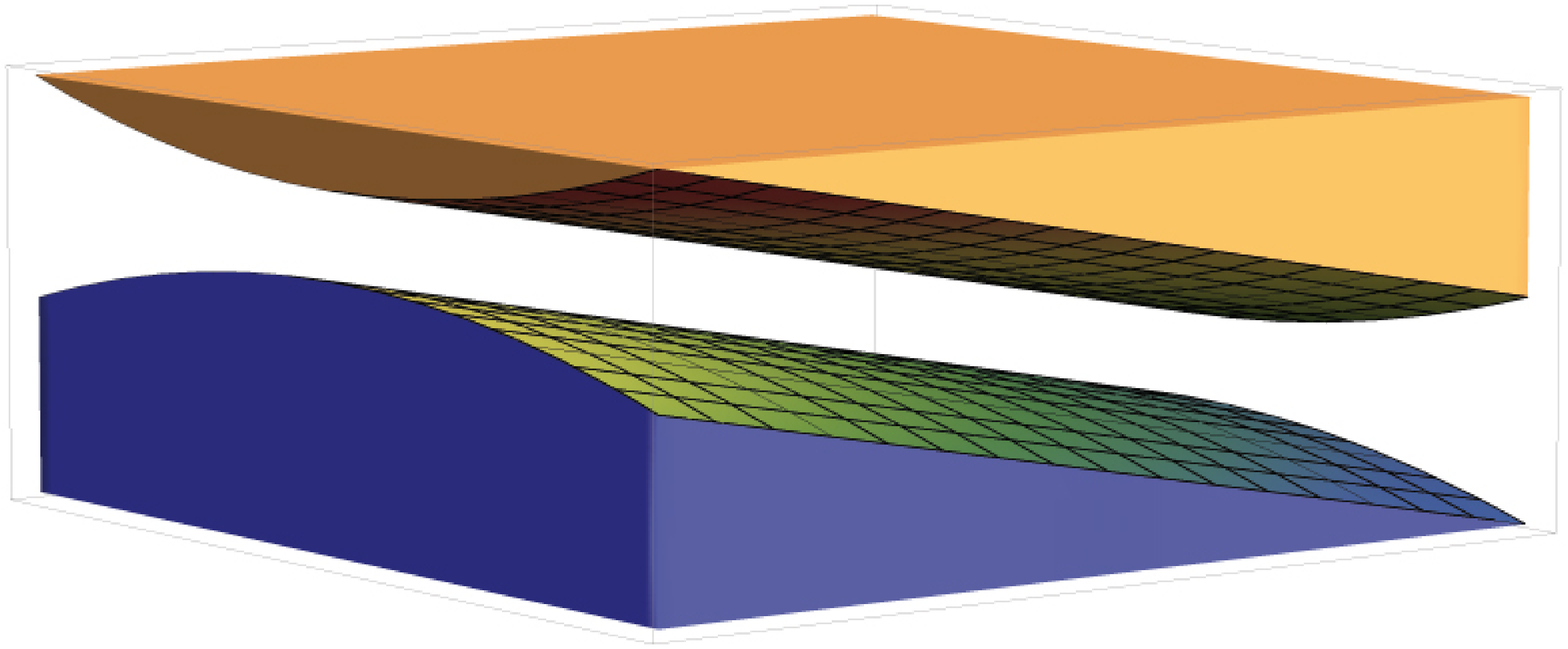} 
  \end{minipage}
  \caption{\textit{Left}: Relativistic dispersion relation for vanishing electric field. 
  \textit{Right}: The electric field leads to a tilt in the relativistic dispersion relation such that tunneling out of the filled Dirac sea to the continuum is possible.}
  \label{tiltedpotential}
\end{center}
\end{figure}

In order to create electron-positron pairs out of the vacuum, it is necessary to overcome the threshold energy of $2mc^2$ from the filled Dirac sea to the unoccupied energy levels. Basically, there are two mechanism to achieve this: 
\begin{itemize}
  \item{\textit{Schwinger pair creation} \cite{Schwinger:1951nm,Heisenberg:1935qt}: Virtual
electron-positron fluctuations may gain the necessary energy $e\mathcal{E}L>2mc^2$ by acceleration over a distance $L$ in a static or slowly varying electric field $\mathcal{E}$. This effect might be considered as tunneling effect and is as such exponentially suppressed.}
 \item{\textit{Dynamical pair creation} \cite{Brezin:1970xf,Marinov:1977gq}: In a time-dependent electric field with characteristic frequency $\omega$, the threshold energy might be provided by a single photon $\hbar\omega>2mc^2$. For $\hbar\omega<2mc^2$, this effect might still occur by absorbing a multiple number of photons, but then becomes strongly suppressed as well.}
\end{itemize}
An important parameter facilitating this characterization is the Keldysh parameter, first introduced in the context of atomic ionization \cite{Keldysh:1964ud} but later also taken over in the context of electron-positron pair creation from vacuum \cite{Brezin:1970xf}:
\begin{equation}
	\gamma = \frac{\tau_T}{\tau}=\frac{m}{e\mathcal{E}\tau} \; . \label{keldysh}
\end{equation}
Here, $\tau_T=m/e\mathcal{E}$ is the characteristic time of a tunneling event. On the other hand, $\tau$ denotes the characteristic time scale of the applied electric field. The ability to tunnel is facilitated by the electric field $\mathcal{E}$ which tilts the potential around the electric field axis, as shown in Fig.~\ref{tiltedpotential}. 

Consequently, the Keldysh parameter discriminates between two regimes:
\begin{itemize}
  \item{\textit{Adiabatic regime} $\gamma\ll 1$ ($\tau_T\ll\tau$): Tunneling through the barrier is possible since the electric field might be considered as static at any instant of time. In this regime, electron-positron pairs are created spontaneously by means of a non-perturbative tunneling process.} 
  \item{\textit{Anti-adiabatic regime} $\gamma\gg 1$ ($\tau_T\gg\tau$): Tunneling through the barrier is not possible because the shape of the electric field changes much faster than the time needed to tunnel. In this perturbative regime, electron-positron pairs might only be created by absorption of the necessary threshold energy and not via a tunneling event.}
\end{itemize}

\subsection{Electric Field with Two Time Scales}

Multi-photon pair creation has already been experimentally verified at the SLAC E-144 experiment \cite{Bamber:1999zt}, whereas the Schwinger mechanism has not been observed so far due to its strong exponential suppression of the order of $\exp(-\pi \mathcal{E}_c/\mathcal{E})$. Hence, various mechanisms have been proposed in order to partly overcome this suppression of the Schwinger mechanism by taking advantage of pair creation in the multiphoton regime \cite{Schutzhold:2008pz,Dunne:2009gi,DiPiazza:2009py, Bulanov:2010ei}. 

\begin{figure}[t!]
  \centering
  \includegraphics[width=0.48\textwidth]{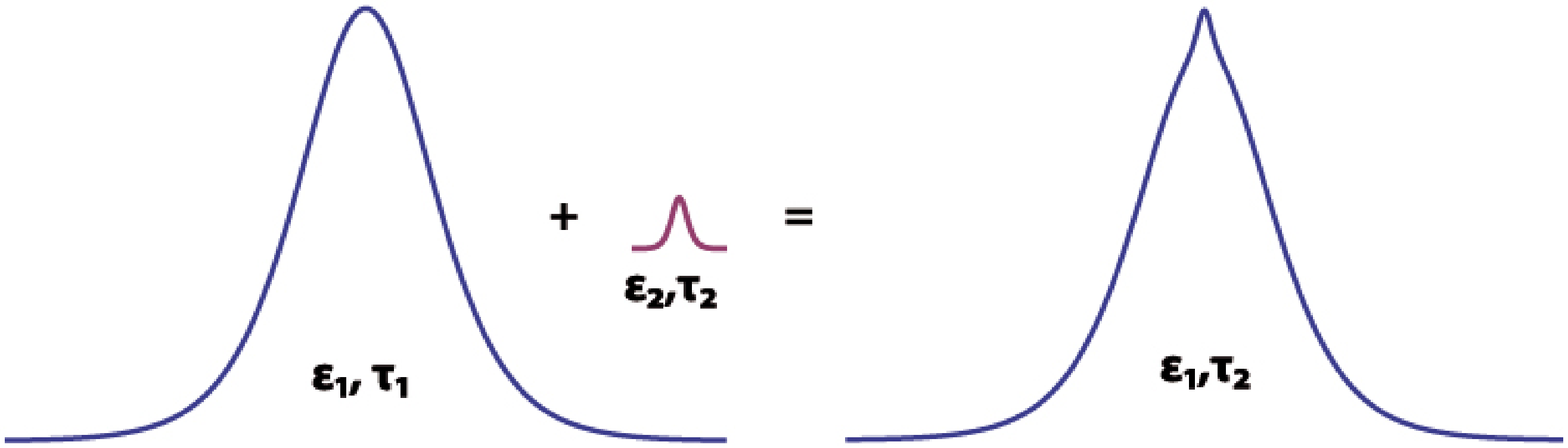}
  \caption{Superposition of two Sauter-type electric fields Eq.~(\ref{sech2}) for $t_0=0$. The 
  relevant scales in this system are the larger field strength $ \mathcal{E}_1 $ and the shorter  time scale $\tau_2$.
  \label{combpic}} 
\end{figure}

The idea of the dynamically assisted Schwinger mechanism is rather simple \cite{Schutzhold:2008pz}: Superimposing a strong but slowly varying electric field with a weak but rapidly varying electric field leads to a decrease of the tunneling barrier and, consequently, to a drastic enhancement of the electron-positron pair creation rate. As a very simple example we consider two Sauter-type electric fields
\footnote{Note, that due to the fact that the Dirac equation is exactly solvable in terms of hypergeometric functions in a single background field Eq.~(\ref{sech}), we are able to find an analytic expression for the distribution function $f(\mathbf{q},t)$ at both finite and asymptotic times \cite{Hebenstreit:2010vz}. For simplicity, we only give the result of the asymptotic momentum distribution here:
\begin{equation}
  \label{dist_sech}
  f(\mathbf{q},\infty)=\frac{2\sinh\left(\pi\left[e\mathcal{E}\tau^2+\mu-\nu\right]\right)\sinh\left(\pi\left[e\mathcal{E}\tau^2-\mu+\nu\right]\right)}{\sinh(2\pi\mu)\sinh(2\pi\nu)} \ , 
\end{equation}
with 
\begin{eqnarray}
  \mu&=&\frac{\tau}{2}\sqrt{m^2+\mathbf{q}_\perp^2+(q_3-e\mathcal{E}\tau)} \ , \nonumber\\  
  \nu&=&\frac{\tau}{2}\sqrt{m^2+\mathbf{q}_\perp^2+(q_3+e\mathcal{E}\tau)} \ . \nonumber
\end{eqnarray}}
\begin{equation}
  \label{sech}
  E_i(t) = \mathcal{E}_i\operatorname{sech}^2\left(t/\tau_i\right)\, . 
\end{equation}
In our calculations we take $E_1(t)$ as the strong field, whereas $E_2(t)$ is considered as the weak field. Consequently, we have $\mathcal{E}_2\ll \mathcal{E}_1$ and $\tau_2\ll\tau_1$. In terms of the Keldysh parameter we have $\gamma_1\ll1$ and $\gamma_2\gg1$. We then consider the superposition of two such electric fields (cf. Fig.~\ref{combpic}):
\begin{equation}
  \label{sech2}
E(t)=\mathcal{E}_1\operatorname{sech}^2\left(t/\tau_1\right)+\mathcal{E}_2\operatorname{sech}^2\left((t-t_0)/\tau_2\right) \ . 
\end{equation}
The parameter $t_0$ accounts for a possible shift of the individual pulse maxima with respect to each other. 

It must be stressed that the relevant Keldysh parameter for this problem is neither $\gamma_1$ nor $\gamma_2$ but a combination of the dominant scales of the two different electric pulses \cite{Schutzhold:2008pz}. These relevant scales are given by the larger field strength $\mathcal{E}_1 $ and the shorter time scale $\tau_2$:
\begin{equation}
  \gamma=\frac{m}{e\mathcal{E}_1\tau_2} \ .
\end{equation}
This combination of different scales results in a subtle change in the pair production process by means of a non-trivial combination of Schwinger and dynamical pair creation. Consequently, the resulting distribution function $f(\mathbf{q},t)$ will not be given by a naive addition of $f_1(\mathbf{q},t)$ for the Schwinger mechanism and $f_2(\mathbf{q},t)$ for dynamical pair creation, as we will see in the next section. In our calculations we consider a given strong electric field $E_1(t)$ in the adiabatic regime  and a weak electric field $E_2(t)$ which can be varied by changing either the field strength $\mathcal{E}_2$ or the pulse length $\tau_2$.

\begin{figure}[h!]
  \includegraphics[width=0.47\textwidth]{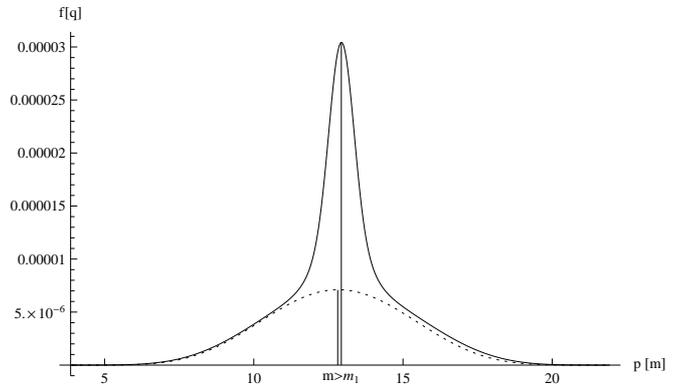}
  \caption{Comparison of the distribution function $f(\mathbf{q},\infty)$ (solid line) with
  $f_1(\mathbf{q},\infty)$ (dotted line). The positions of the maxima of the distribution 
  functions are marked by $m$ and $m_1$, respectively. The parameters are given in the text. $f_2(\mathbf{q},\infty)$ is not shown because it is orders of magnitude smaller.
  \label{combt0}}
\end{figure}

\section{Numerical Results}
\label{sec3}

We now turn to the numerical results for the dynamically assisted Schwinger mechanism in an electric field with multiple scales. In a first instance we focus on the asymptotic distribution function $f(\mathbf{q},\infty)$ of created particles. Subsequently, we discuss the enhancement in the number density $n[e^+e^-]$ of created particles in a specific parameter region due to the non-trivial combination of Schwinger and dynamical pair creation.

\subsection{Momentum Spectrum}

In Fig.~\ref{combt0} we consider the dynamically assisted Schwinger mechanism for a given setup, with the parameters for the strong pulse $\{\mathcal{E}_1=0.25\mathcal{E}_c,\tau_1=10^{-4}\,eV^{-1}\}$ and the weak pulse $\{\mathcal{E}_2=0.025\mathcal{E}_c,\tau_2= 7\cdot10^{-6}\,eV^{-1} \}$. For these parameters, $\gamma_1 \approx 0.078 $ is deep in the non-perturbative regime whereas $\gamma_2\approx 11.18$ is well in the multi-photon region. In this figure, we compare the momentum spectrum for a single Schwinger-like pulse with the momentum spectrum for a combined pulse Eq.~(\ref{sech2}) for $t_0=0$. We should first make two remarks:
\begin{itemize}
  \item{The parameter sets $\{\mathcal{E}_1,\tau_1\}$ and $\{\mathcal{E}_2,\tau_2\}$ are chosen such that the contribution due to Schwinger pair production $f_1(\mathbf{q},\infty)$ is orders of magnitude larger than that due to dynamical pair production $f_2(\mathbf{q},\infty )$. In order to observe the dynamically assisted Schwinger mechanism, it is of great importance to consider a parameter region which is in this sense dominated non-perturbatively (Schwinger-like).}
  \item{The distribution function for single pulses is easily found by the analytic formula Eq.~(\ref{dist_sech}). This formula also served as a numerics benchmark throughout this work.}
\end{itemize}

We observe, that the momentum spectrum for the combined pulse $f(\mathbf{q},\infty)$ is \textit{not} given by naively adding $f_1(\mathbf{q},\infty)$ and $f_2(\mathbf{q},\infty)$ but shows a strong non-linear behavior. In fact, $f(\mathbf{q},\infty)$ and $f_1(\mathbf{q},\infty)$ nearly coincide for small kinetic momenta but differ for large kinetic momenta $f(\mathbf{q}, \infty)>f_1(\mathbf{q},\infty)$. Additionally, the momentum spectrum is strongly enhanced around the maximum of $f_1(\mathbf{q},\infty)$ (denoted as $m_1$) due to the influence of the short pulse $\{\mathcal{E}_2,\tau_2\}$. Moreover, the maximum of $f(\mathbf{q},\infty)$ (denoted as $m$) is shifted a little towards higher momenta compared to the pure Schwinger mechanism $f_1(\mathbf{q}, \infty)$, i.e. $m>m_1$. 
\begin{figure}[t!]
  \centering
    \includegraphics[width=0.45\textwidth]{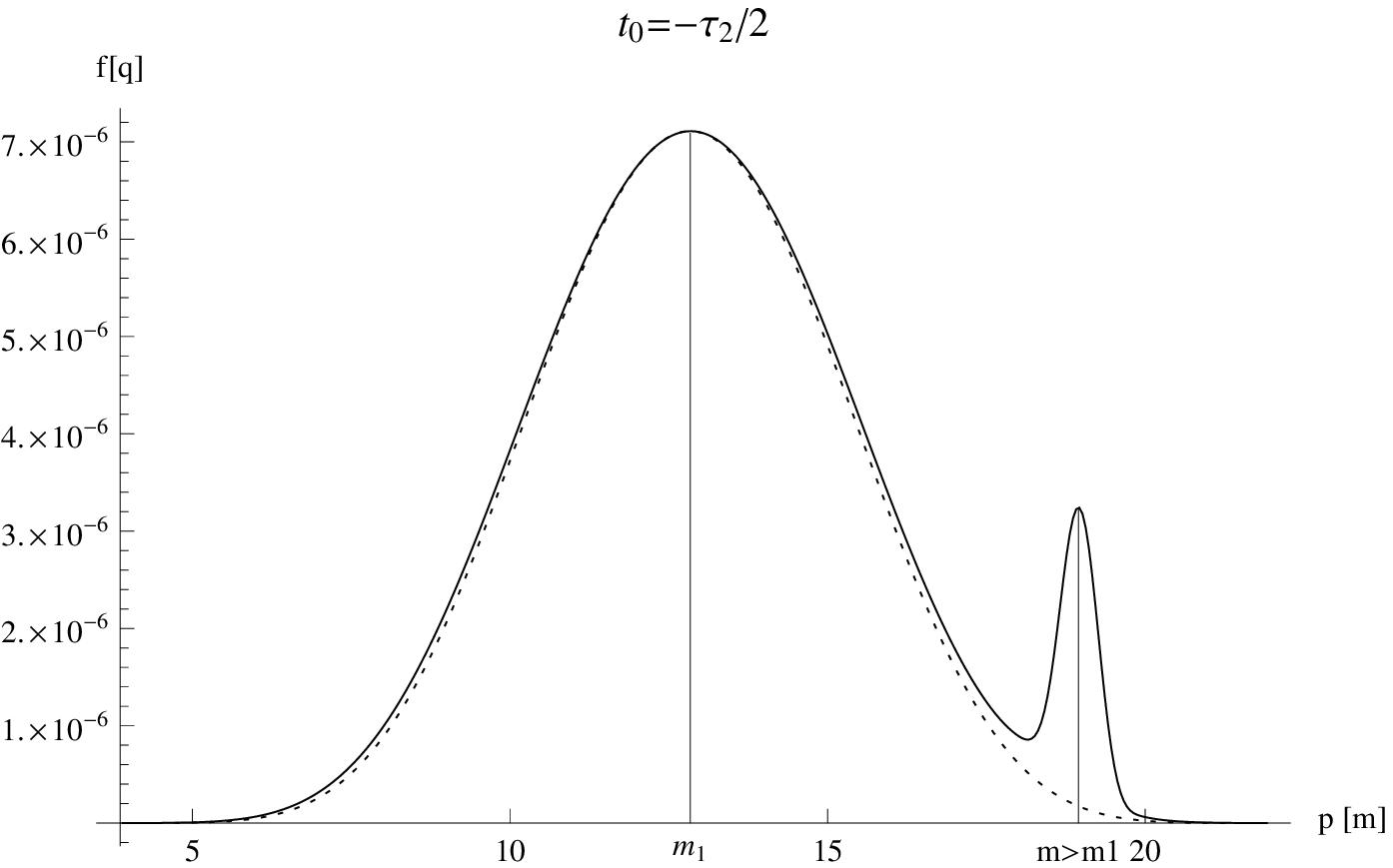}
    \vskip 0.19cm 
    \includegraphics[width=0.45\textwidth]{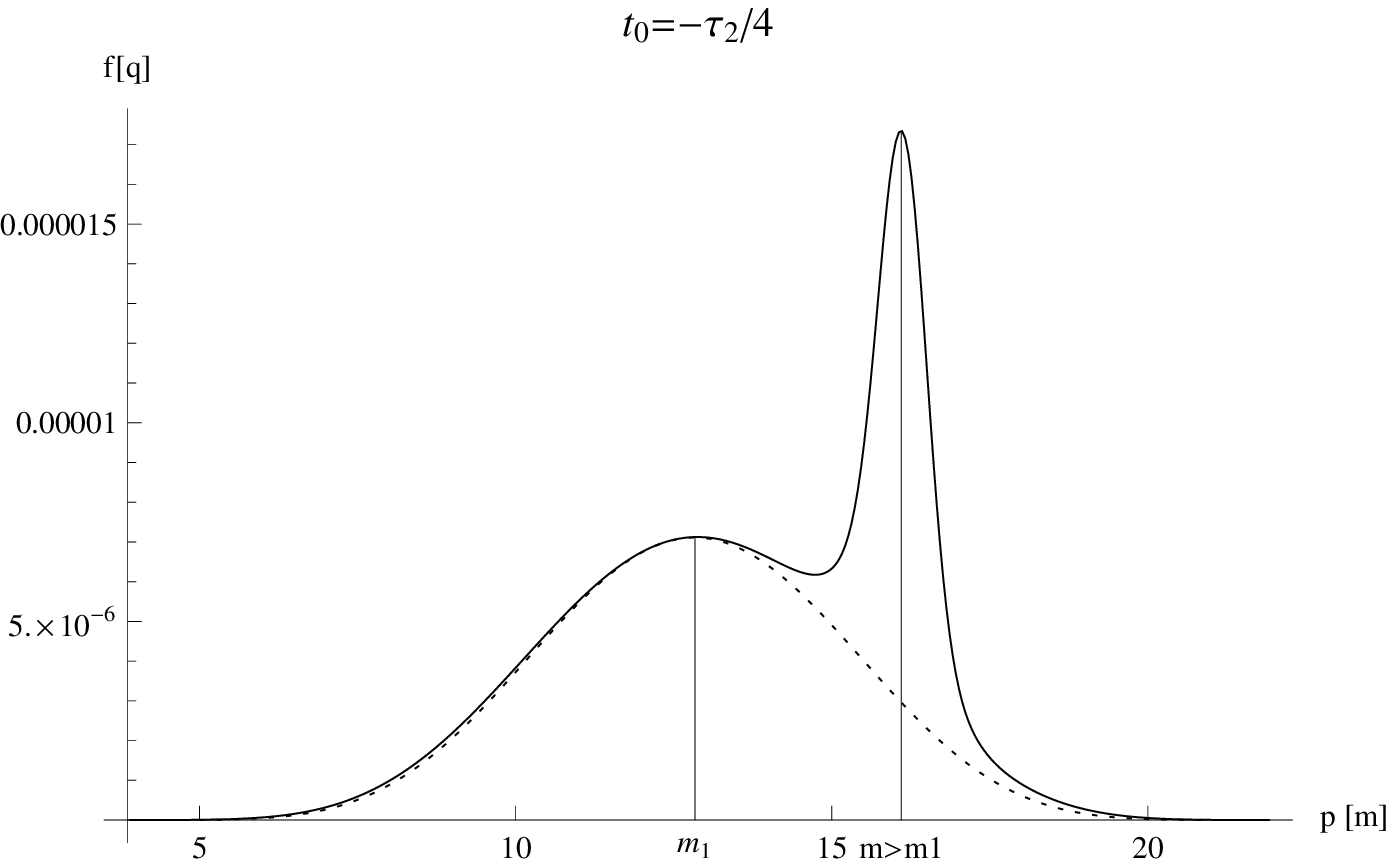} 
    \vskip 0.19cm
    \includegraphics[width=0.45\textwidth]{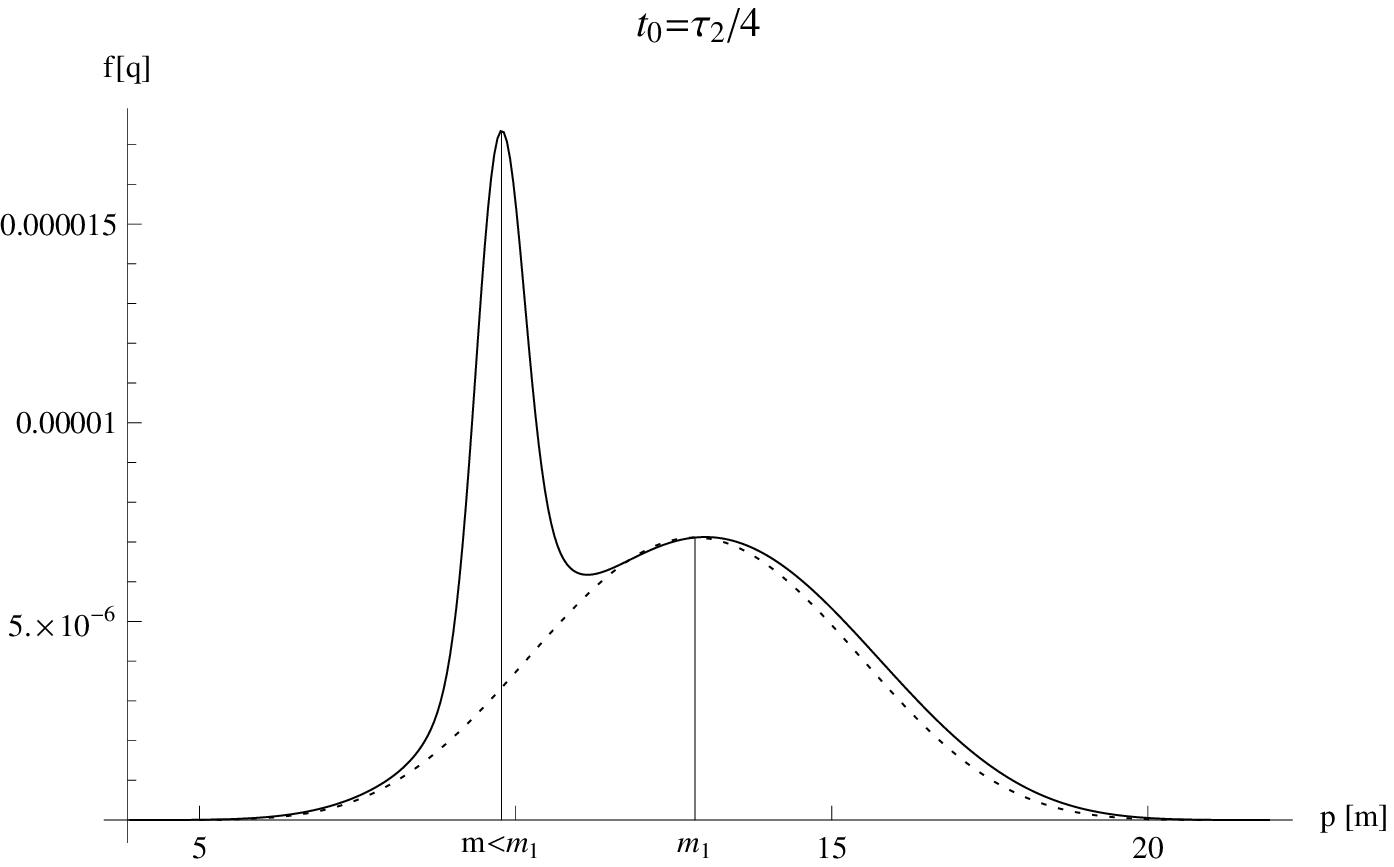} 
    \vskip 0.19cm
    \includegraphics[width=0.45\textwidth]{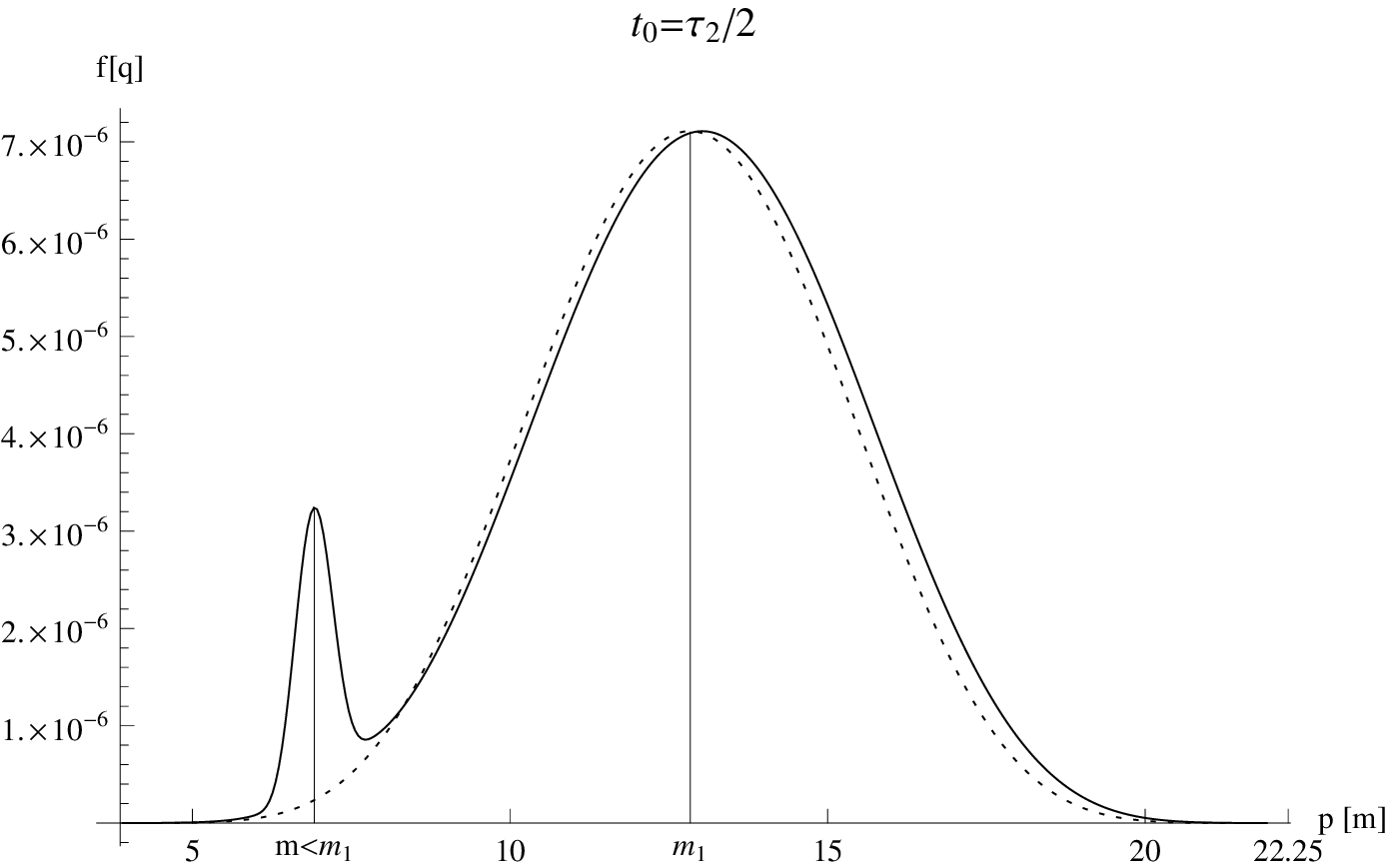}
    \vskip 0.19cm
  \caption{Comparison of the distribution function $f(\mathbf{q},\infty)$ (solid line) with
  $f_1(\mathbf{q},\infty)$ (dotted line). Depending on $t_0$, the maximum of the weak pulse 
  occurs either prior or after the maximum of the strong pulse. The other parameters are chosen as in Fig.~\ref{combt0}.
  \label{t0_dep}}  
\end{figure}

In order to better understand this behavior, let us consider the influence of different $t_0$ values on the pair creation process in Fig.~\ref{t0_dep}. For $t_0<0$ (two upper figures), i.e. the maximum of the weak pulse occurs \textit{prior} to the maximum of the strong pulse, we observe that the two curves coincide around $m_1$. Due to the fact that the weak pulse acts prior to the strong pulse, this field configuration behaves much like a single strong pulse. The only difference is that particles which are created by the weak pulse are accelerated by the subsequent strong pulse such that $m>m_1$. Oppositely for $t_0>0$ (two lower figures), i.e. the maximum of the weak pulse occurs \textit{after} the maximum of the strong pulse. In this case, the particles which are created by the strong pulse get an extra kick by the subsequent weak pulse such that $f(\mathbf{q},\infty)$ is shifted to larger momenta compared to $f_1(\mathbf{q},\infty)$. Additionally, the pairs which are created by the weak pulse are accelerated less than in the previous case such that $m<m_1$.

For $t_0=0$ we observe a combination of those two effects (cf. Fig.~\ref{combt0}): On the one hand we find that $f(\mathbf{q},\infty)>f_1(\mathbf{q},\infty)$ for large momenta, i.e. the particles get an extra kick by the weak pulse. On the other hand, we also find that $m>m_1$, i.e. the particles which are created by the weak pulse are still strongly accelerated by the strong pulse. 

\begin{figure}[b!]
  \centering
  \includegraphics[width=0.47\textwidth]{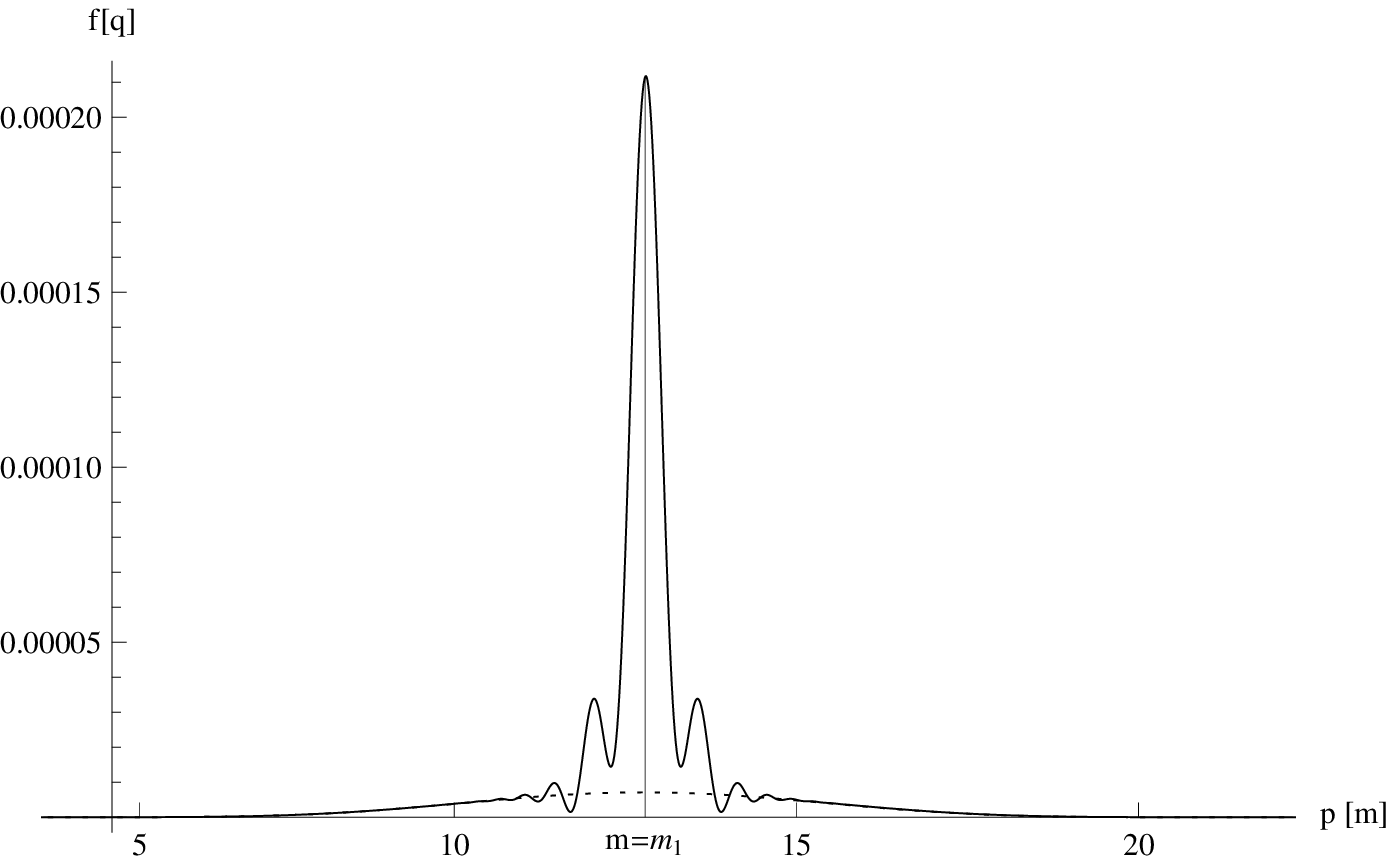}
  \includegraphics[width=0.47\textwidth]{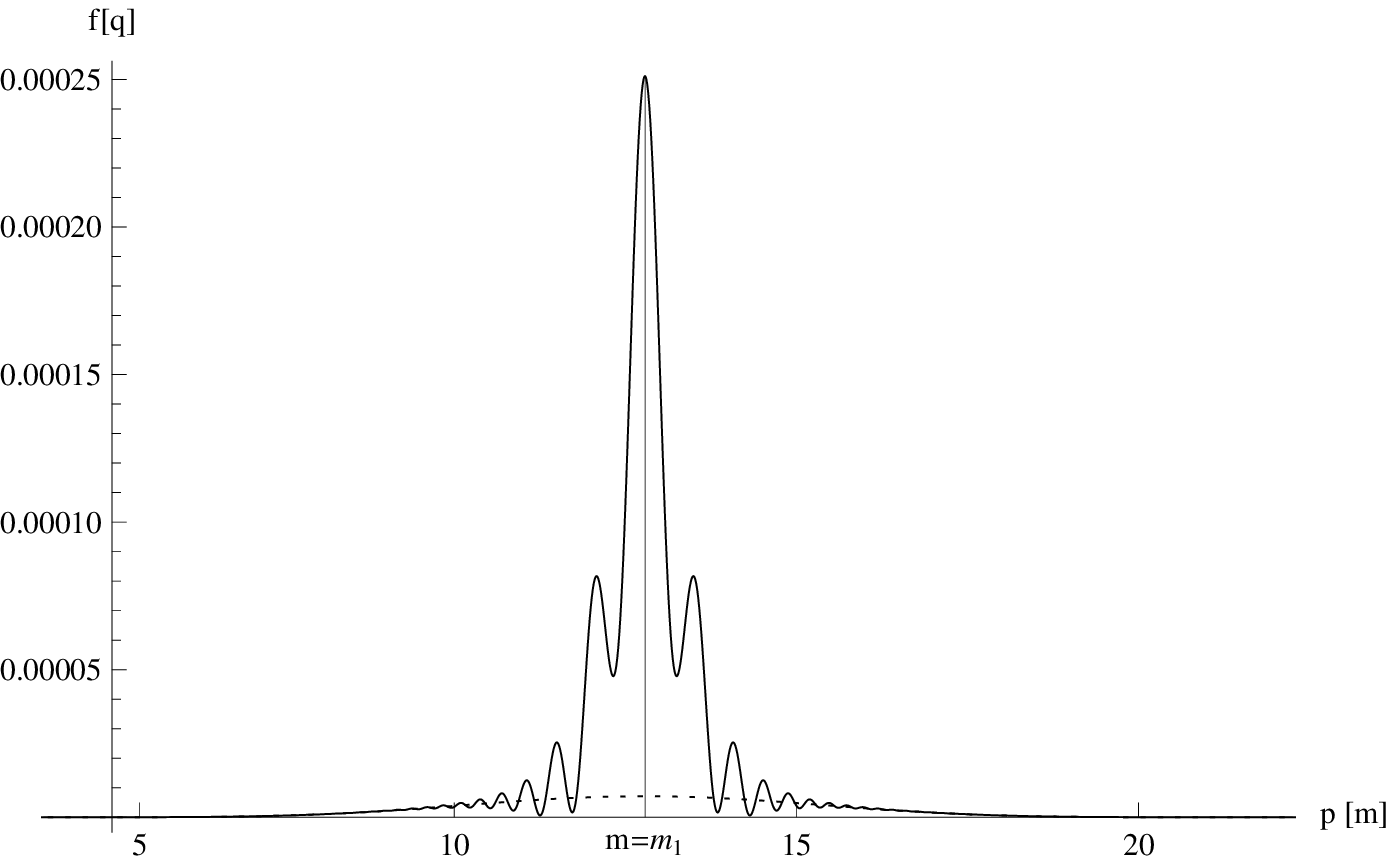}
  \caption{Comparison of the distribution function $f(\mathbf{q},\infty)$ (solid line) with
  $f_1(\mathbf{q},\infty)$ (dotted line) for different values of $\tau_2$. \textit{Top:} 
  $\tau_2=2\cdot10^{-6}\,eV^{-1} $\textit{Bottom:} $\tau_2=10^{-6}\,eV^{-1}$ The other parameters are 
  chosen as in Fig.~\ref{combt0}. 
  \label{combwiggles}}
\end{figure}

Due to the fact that we obtain the maximal enhancement at $t_0=0$, we consider this situation in the following. Before discussing the enhancement of the number density $n[e^+e^-]$ in the next section, we want to investigate the momentum spectrum depending on the pulse length $\tau_2$ of the weak pulse in Fig.~\ref{combwiggles}. 

The weak pulse becomes shorter by decreasing the value of $\tau_2$, i.e., the energy of the corresponding quanta gets larger. Accordingly, it is getting easier to create an electron-positron pair via dynamical pair creation. As soon as the value of $\tau_2\lesssim 5\cdot10^{-6}\,eV^{-1}$, we observe distinct momentum signatures around the maximum $m$ in tune with the inverse time scale $1/\tau_2$. This might be interpreted as the onset of multi-photon domination, i.e., whilst the dynamically assisted Schwinger mechanism was still non-perturbatively dominated in Fig.~\ref{combt0}, the effect becomes mostly perturbative in Fig.~\ref{combwiggles}. 

\subsection{Number Density}

\begin{figure}[t!]
  \centering
  \includegraphics[width=0.47\textwidth]{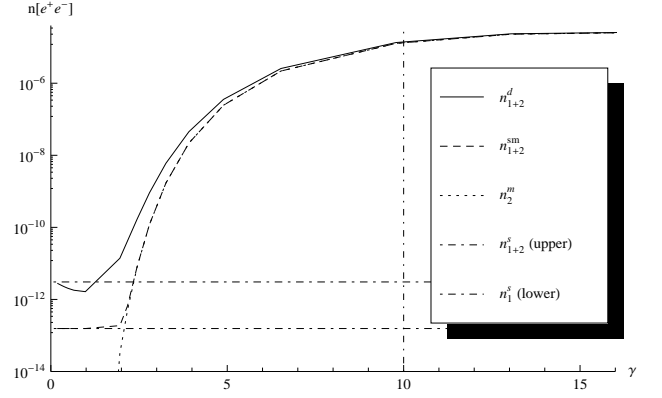}
  \caption{Comparison of the number density $n[e^+e^-]$ according to the various effects. The parameters are given in the text.
  \label{combnum}}
\end{figure}

We now turn to the investigation of the number density $n[e^+e^-]$ for the dynamically assisted Schwinger mechanism. In Fig.~\ref{combnum} we vary the pulse length $\tau_2$ of the weak pulse and, accordingly, the combined Keldysh parameter $\gamma$. The other parameters are chosen to be $\{\mathcal{E}_1=0.1\mathcal{E}_c,\tau_1=2\cdot 10^{-4} eV^{-1} \}$ for the strong pulse (this corresponds to $\gamma_1\sim0.097$) and $\{\mathcal{E}_2=0.01\mathcal{E}_c,\tau_2\ \textit{variable} \}$ for the weak pulse:
\begin{itemize}
  \item The number density $n^{s}_1[e^+e^-]$ of a single strong pulse $\{\mathcal{E}_1,\tau_1\}$ (\textit{pure Schwinger mechanism}) is represented by the \textit{lower lying dashed-dotted line}.
  \item The number density $n^{s}_{1+2}[e^+e^-]$ of a single strong pulse $\{\mathcal{E}_1+ \mathcal{E}_2,\tau_1\}$ (\textit{pure Schwinger mechanism}) is represented by the \textit{higher lying dashed-dotted line}.
  \item The number density $n^{m}_2[e^+e^-]$ of a single weak pulse $\{\mathcal{E}_2,\tau_2\}$ (\textit{pure multi-photon effect}) with $\tau_2$ being varied is represented by the \textit{dotted line}.
   \item The number density $n^{sm}_{1+2}[e^+e^-]$ which is obtained by naively adding the contributions of both the Schwinger and the multi-photon effect $n^{s}_1[e^+e^-]+n^{m}_2[e^+e^-]$ is represented by the \textit{dashed line}.
  \item The number density $n^{d}_{1+2}[e^+e^-]$ which is obtained by our numerical calculation of the \textit{dynamically assisted Schwinger mechanism} is represented by the \textit{solid line}.
\end{itemize}
Additionally we have a dotted vertical line which corresponds to the parameter set at which the weak pulse has a time scale being equal to the Compton time of the electron, i.e., $\tau_2=1/m$. Note that these calculations have been done for $q_\perp=0$, i.e. ignoring the perpendicular momentum. Including the orthogonal momentum in the calculations is straightforward but not done for computational reasons. We observe the following behavior:

\begin{itemize}
  \item{For a very small Keldysh parameter $\gamma\ll1$ such that $ \tau_2 = \tau_1 $, we obtain the same result as the pure Schwinger mechanism with the parameters $\{\mathcal{E}_1+\mathcal{E}_2,\tau_1\}$. This is exactly what we expect since:
  \begin{equation} 
    E(t)=E_1(t)+E_2(t)=(\mathcal{E}_1+\mathcal{E}_2)\operatorname{sech}^2\left(t/\tau_1\right)
  \end{equation}}
  \item{For a very large Keldysh parameter $\gamma\gg1$ the multiple time scale effect asymptotically coincides with the multi-photon result. This is reasonable because for very large $\gamma$ the multi-photon effect dominates the Schwinger mechanism by orders of magnitude.}
  \item{In between these two extreme cases we have the region which is of special interest to us. Here, we see clear signatures of the dynamically assisted Schwinger mechanism: For $1\lesssim\gamma \lesssim2$ we observe that $n^{sm}_{1+2}[e^+e^-]$, i.e. the naively added Schwinger and multi-photon effect, is well below $n^{d}_{1+2}[e^+e^-]$, i.e. the actual outcome of the dynamically assisted Schwinger mechanism.}
\end{itemize}

\begin{figure}[t!]
  \centering
  \includegraphics[width=0.47\textwidth]{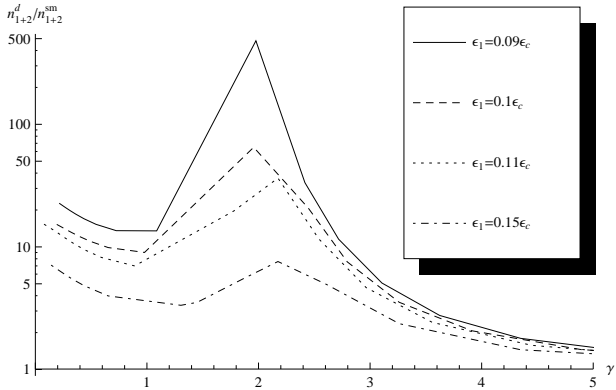}
  \caption{Relative Enhancement of the number density $n^{d}_{1+2}[e^+e^-]/n^{sm}_{1+2 }[e^+e^-]$ due to the dynamically assisted Schwinger mechanism. The parameters are given in the text.
  \label{combcomp}}
\end{figure}

Actually, the relative enhancement of the number density $n^{d}_{1+2}[e^+e^-]/n^{sm}_{1+2 }[e^+e^-]$ strongly depends on the various parameters, as shown in Fig.~\ref{combcomp}. To this end, we vary $\mathcal{E}_1$ but keep the ratio between the field strengths $\mathcal{E}_2=0.1\mathcal{E}_1$ constant. Additionally, $\tau_1 = 2 \cdot 10^{-4} eV^{-1}$ whereas $\tau_2$ is varied again.

The most important result is that in a region around $\gamma\sim2$ the
relative enhancement reaches a maximum with an enhancement factor of up to $500$
for a given parameter set.

\section{Conclusions and Outlook}

We have presented a detailed calculation for an example of the dynamically
assisted Schwinger mechanism which has been suggested recently
\cite{Schutzhold:2008pz,Dunne:2009gi}. This mechanism, i.e., electron-positron
pair production from vacuum by a combination of laser pulses with different time
scales, suggests that the rate of produced pairs can be significantly enhanced
by dynamical effects. First of all, we verified previous results, most
prominently that one encounters a region around the Keldysh parameter
$\gamma\sim2$ where the relative enhancement reaches a maximum.  Hereby,
enhancement factors up to $500$ have been obtained. Second, whereas the former
investigations \cite{Schutzhold:2008pz,Dunne:2009gi} used the world-line
instanton approach,  the presented calculations are carried out within the
framework of  quantum kinetic theory. This method intrinsically enables us to
additionally  present valuable momentum space information of the generated
positron spectrum.

On the basis of these results we conclude that the search for the dynamically
assisted Schwinger mechanism could be facilitated by an ideal parameter choice in
future experiments.  It is obvious that the  trivial combination of the two 
electric pulses considered here is neither optimal nor accomplishable
experimentally.  Therefore, a natural next step will be the investigation of
more realistic electric pulses tuning the parameters for an increased pair
production. Such pulse shaping in order to obtain an enhancement factor close to
the greatest possible one 
 even for more realistic electric field configurations is,
although complicated, certainly feasible. Given  the recent progress in theory
and experiment a  measurement of the Schwinger mechanism in this decade seems
realistic.

\section*{Acknowledgments}
We thank Gerald Dunne and Holger Gies for helpful discussions and a critical
reading of the manuscript.\\
{ FH 
is supported by the DOC program of the Austrian Academy of 
Sciences and by the FWF doctoral program DK-W1203.}

%\bibliographystyle{elsbiblio.bst}
%\bibliography{bibliography}

\begin{thebibliography}{100}

%\cite{Schwinger:1951nm}
\bibitem{Schwinger:1951nm}
  J.~S.~Schwinger,
  %``On gauge invariance and vacuum polarization,''
  Phys.\ Rev.\  {\bf 82} (1951) 664.
  %%CITATION = PHRVA,82,664;%%

%\cite{Sauter:1931zz}
\bibitem{Sauter:1931zz}
  F.~Sauter,
  %``Uber das Verhalten eines Elektrons im homogenen elektrischen Feld nach der
  %relativistischen Theorie Diracs,''
  Z.\ Phys.\  {\bf 69} (1931) 742.
  %%CITATION = ZEPYA,69,742;%%

%\cite{Heisenberg:1935qt}
\bibitem{Heisenberg:1935qt}
  W.~Heisenberg and H.~Euler,
  %``Consequences of Dirac's theory of positrons,''
  Z.\ Phys.\  {\bf 98} (1936) 714.
%  [arXiv:physics/0605038].
  %%CITATION = ZEPYA,98,714;%%

%\cite{Alkofer:2001ik}
\bibitem{Alkofer:2001ik}
  R.~Alkofer {\it et al.},%, M.~B.~Hecht, C.~D.~Roberts, S.~M.~Schmidt and D.~V.~Vinnik,
  %``Pair Creation and an X-ray Free Electron Laser,''
  Phys.\ Rev.\ Lett.\  {\bf 87} (2001) 193902.
  %[arXiv:nucl-th/0108046].
  %%CITATION = PRLTA,87,193902;%%

%\cite{Ringwald:2001ib}
\bibitem{Ringwald:2001ib}
  A.~Ringwald,
  %``Pair production from vacuum at the focus of an X-ray free electron
  %laser,''
  Phys.\ Lett.\  B {\bf 510} (2001) 107.
  %[arXiv:hep-ph/0103185].
  %%CITATION = PHLTA,B510,107;%%

%\cite{Blaschke:2005hs}
\bibitem{Blaschke:2005hs}
  D.~B.~Blaschke {\it et al.},% A.~V.~Prozorkevich, C.~D.~Roberts, S.~M.~Schmidt and S.~A.~Smolyansky,
  %``Pair production and optical lasers,''
  Phys.\ Rev.\ Lett.\  {\bf 96}, (2006) 140402.
  %[arXiv:nucl-th/0511085].
  %%CITATION = PRLTA,96,140402;%%

%\cite{Schutzhold:2008pz}
\bibitem{Schutzhold:2008pz}
  R.~Schutzhold, H.~Gies and G.~Dunne,
  %``Dynamically assisted Schwinger mechanism,''
  Phys.\ Rev.\ Lett.\  {\bf 101} (2008) 130404.
%  [arXiv:0807.0754 [hep-th]].
  %%CITATION = PRLTA,101,130404;%%

%\cite{Dunne:2009gi}
\bibitem{Dunne:2009gi}
  G.~V.~Dunne, H.~Gies and R.~Schutzhold,
  %``Catalysis of Schwinger Vacuum Pair Production,''
  Phys.\ Rev.\  D {\bf 80} (2009) 111301.
%  [arXiv:0908.0948 [hep-ph]].
  %%CITATION = PHRVA,D80,111301;%%

%\cite{Hebenstreit:2009km}
\bibitem{Hebenstreit:2009km}
  F.~Hebenstreit, R.~Alkofer, G.~V.~Dunne and H.~Gies,
  %``Momentum signatures for Schwinger pair production in short laser pulses
  %with sub-cycle structure,''
  Phys.\ Rev.\ Lett.\  {\bf 102} (2009) 150404.
  %[arXiv:0901.2631 [hep-ph]].
  %%CITATION = PRLTA,102,150404;%%

%\cite{Hebenstreit:2008ae}
\bibitem{Hebenstreit:2008ae}
  F.~Hebenstreit, R.~Alkofer and H.~Gies,
  %``Pair Production Beyond the Schwinger Formula in Time-Dependent Electric
  %Fields,''
  Phys.\ Rev.\  D {\bf 78} (2008) 061701.
  %%CITATION = PHRVA,D78,061701;%%

%\cite{Ruf:2009zz}
\bibitem{Ruf:2009zz}
  M.~Ruf {\it et al.},% G.~R.~Mocken, C.~Muller, K.~Z.~Hatsagortsyan and C.~H.~Keitel,
  %``Pair production in laser fields oscillating in space and time,''
  Phys.\ Rev.\ Lett.\  {\bf 102} (2009) 080402.
  %[arXiv:0810.4047 [physics.atom-ph]].
  %%CITATION = PRLTA,102,080402;%%

%\cite{DiPiazza:2009py}
\bibitem{DiPiazza:2009py}
  A.~Di Piazza, E.~Lotstedt, A.~I.~Milstein and C.~H.~Keitel,
  %``Barrier control in tunneling e^+ e^- photoproduction,''
  Phys.\ Rev.\ Lett.\  {\bf 103} (2009) 170403.
%  [arXiv:0906.0726 [hep-ph]].
  %%CITATION = PRLTA,103,170403;%%

%\cite{Bulanov:2010ei}
\bibitem{Bulanov:2010ei}
  S.~S.~Bulanov {\it et al.},% V.~D.~Mur, N.~B.~Narozhny, J.~Nees and V.~S.~Popov,
  %``Multiple colliding electromagnetic pulses: a way to lower the threshold of
  %$e^+e^-$ pair production from vacuum,''
  Phys.\ Rev.\ Lett.\  {\bf 104} (2010) 220404.
%  [arXiv:1003.2623 [hep-ph]].
  %%CITATION = PRLTA,104,220404;%%

\bibitem{Orthaber:2010}
  M.~Orthaber, Diploma thesis, University of Graz, 2010. [\url{http://physik.uni-graz.at/~mao/}]

%\cite{Bell:2008zzb}
\bibitem{Bell:2008zzb}
  A.~R.~Bell and J.~G.~Kirk,
  %``Possibility of Prolific Pair Production with High-Power Lasers,''
  Phys.\ Rev.\ Lett.\  {\bf 101} (2008) 200403.
  %%CITATION = PRLTA,101,200403;%%

%\cite{Fedotov:2010ja}
\bibitem{Fedotov:2010ja}
  A.~M.~Fedotov, N.~B.~Narozhny, G.~Mourou and G.~Korn,
  %``Limitations on the attainable intensity of high power lasers,''
  Phys.\ Rev.\ Lett.\  {\bf 105} (2010) 080402.
%  [arXiv:1004.5398 [hep-ph]].
  %%CITATION = PRLTA,105,080402;%%

%\cite{Bulanov:2010gb}
\bibitem{Bulanov:2010gb}
  S.~S.~Bulanov {\it et al.},% T.~Z.~Esirkepov, A.~G.~R.~Thomas, J.~K.~Koga and S.~V.~Bulanov,
  %``On the Schwinger limit attainability with extreme power lasers,''
  arXiv:1007.4306 [physics.plasm-ph].
  %%CITATION = ARXIV:1007.4306;%%

%\cite{Kim:2003qp}
\bibitem{Kim:2003qp}
  S.~P.~Kim and D.~N.~Page,
  %``Schwinger pair production in electric and magnetic fields,''
  Phys.\ Rev.\  D {\bf 73} (2006) 065020.
%  [arXiv:hep-th/0301132].
  %%CITATION = PHRVA,D73,065020;%%

%\cite{Tanji:2008ku}
\bibitem{Tanji:2008ku}
  N.~Tanji,
  %``Dynamical view of pair creation in uniform electric and magnetic fields,''
  Annals Phys.\  {\bf 324} (2009) 1691.
%  [arXiv:0810.4429 [hep-ph]].
  %%CITATION = APNYA,324,1691;%%

%\cite{Gies:2005bz}
\bibitem{Gies:2005bz}
  H.~Gies and K.~Klingmuller,
  %``Pair production in inhomogeneous fields,''
  Phys.\ Rev.\  D {\bf 72} (2005) 065001
  [arXiv:hep-ph/0505099].
  %%CITATION = PHRVA,D72,065001;%%

%\cite{Dunne:2006ur}
\bibitem{Dunne:2006ur}
  G.~V.~Dunne and Q.~h.~Wang,
  %``Multidimensional worldline instantons,''
  Phys.\ Rev.\  D {\bf 74} (2006) 065015.
%  [arXiv:hep-th/0608020].
  %%CITATION = PHRVA,D74,065015;%%

%\cite{Kleinert:2008sj}
\bibitem{Kleinert:2008sj}
  H.~Kleinert, R.~Ruffini and S.~S.~Xue,
  %``Electron-Positron Pair Production in Space- or Time-Dependent Electric
  %Fields,''
  Phys.\ Rev.\  D {\bf 78} (2008) 025011.
%  [arXiv:0807.0909 [hep-th]].
  %%CITATION = PHRVA,D78,025011;%%

%\cite{Hebenstreit:2010vz}
\bibitem{Hebenstreit:2010vz}
  F.~Hebenstreit, R.~Alkofer and H.~Gies,
  %``Schwinger pair production in space- and time-dependent electric fields:
  %Relating the Wigner formalism to quantum kinetic theory,''
  Phys.\ Rev.\  D {\bf 82} (2010) 105026.
%  [arXiv:1007.1099 [hep-ph]].
  %%CITATION = PHRVA,D82,105026;%%

%\cite{Gregori:2010uf}
\bibitem{Gregori:2010uf}
  G.~Gregori {\it et al.},
  %``A proposal for testing subcritical vacuum pair production with high power
  %lasers,''
  High Energy Dens.\ Phys.\  {\bf 6} (2010) 166.
%  [arXiv:1005.3280 [hep-ph]].
  %%CITATION = 00520,6,166;%%

%\cite{Kluger:1998bm}
\bibitem{Kluger:1998bm}
  Y.~Kluger, E.~Mottola and J.~M.~Eisenberg,
  %``The quantum Vlasov equation and its Markov limit,''
  Phys.\ Rev.\  D {\bf 58} (1998) 125015.
%  [arXiv:hep-ph/9803372].
  %%CITATION = PHRVA,D58,125015;%%

%\cite{Smolyansky:1997fc}
\bibitem{Smolyansky:1997fc}
  S.~A.~Smolyansky {\it et al.},% G.~Ropke, S.~M.~Schmidt, D.~Blaschke, V.~D.~Toneev and A.~V.~Prozorkevich,
  %``Dynamical derivation of a quantum kinetic equation for particle  production
  %in the Schwinger mechanism,''
  arXiv:hep-ph/9712377.
  %%CITATION = HEP-PH/9712377;%%

%\cite{Schmidt:1998vi}
\bibitem{Schmidt:1998vi}
  S.~M.~Schmidt {\it et al.},% D.~Blaschke, G.~Ropke, S.~A.~Smolyansky, A.~V.~Prozorkevich and V.~D.~Toneev,
  %``A quantum kinetic equation for particle production in the Schwinger
  %mechanism,''
  Int.\ J.\ Mod.\ Phys.\  E {\bf 7} (1998) 709.
%  [arXiv:hep-ph/9809227].
  %%CITATION = IMPAE,E7,709;%%

%\cite{BialynickiBirula:1991tx}
\bibitem{BialynickiBirula:1991tx}
  I.~Bialynicki-Birula, P.~Gornicki and J.~Rafelski,
  %``Phase Space Structure Of The Dirac Vacuum,''
  Phys.\ Rev.\  D {\bf 44} (1991) 1825.
  %%CITATION = PHRVA,D44,1825;%%

%\cite{Vinnik:2001qd}
\bibitem{Vinnik:2001qd}
  D.~V.~Vinnik {\it et al.},% A.~V.~Prozorkevich, S.~A.~Smolyansky, V.~D.~Toneev, M.~B.~Hecht, C.~D.~Roberts and S.~M.~Schmidt,
  %``Plasma production and thermalisation in a strong field,''
  Eur.\ Phys.\ J.\  C {\bf 22} (2001) 341.
%  [arXiv:nucl-th/0103073].
  %%CITATION = EPHJA,C22,341;%%

%\cite{Bloch:1999eu}
\bibitem{Bloch:1999eu}
  J.~C.~R.~Bloch {\it et al.},% V.~A.~Mizerny, A.~V.~Prozorkevich, C.~D.~Roberts, S.~M.~Schmidt, S.~A.~Smolyansky and D.~V.~Vinnik,
  %``Pair creation: Back-reactions and damping,''
  Phys.\ Rev.\  D {\bf 60} (1999) 116011.
%  [arXiv:nucl-th/9907027].
  %%CITATION = PHRVA,D60,116011;%%

%\cite{Roberts:2000aa}
\bibitem{Roberts:2000aa}
  C.~D.~Roberts and S.~M.~Schmidt,
  %``Dyson-Schwinger equations: Density, temperature and continuum strong
  %QCD,''
  Prog.\ Part.\ Nucl.\ Phys.\  {\bf 45} (2000) S1.
%  [arXiv:nucl-th/0005064].
  %%CITATION = PPNPD,45,S1;%%

%\cite{Brezin:1970xf}
\bibitem{Brezin:1970xf}
  E.~Brezin and C.~Itzykson,
  %``Pair production in vacuum by an alternating field,''
  Phys.\ Rev.\  D {\bf 2} (1970) 1191.
  %%CITATION = PHRVA,D2,1191;%%

%\cite{Marinov:1977gq}
\bibitem{Marinov:1977gq}
V.~S.~Popov,
%``Pair Production in a Variable External Field (Quasiclassical approximation)'',
Sov. Phys. JETP {\bf 34} (1972) 709;
  M.~S.~Marinov and V.~S.~Popov,
  %``Electron-Positron Pair Creation From Vacuum Induced By Variable Electric
  %Field,''
  Fortsch.\ Phys.\  {\bf 25} (1977) 373.
  %%CITATION = FPYKA,25,373;%%


%\cite{Keldysh:1964ud}
\bibitem{Keldysh:1964ud}
  L.~V.~Keldysh,
  %``Diagram technique for nonequilibrium processes,''
  Zh.\ Eksp.\ Teor.\ Fiz.\  {\bf 47} (1964) 1515
  [Sov.\ Phys.\ JETP {\bf 20} (1965) 1018].
  %%CITATION = SPHJA,20,1018;%%

%\cite{Bamber:1999zt}
\bibitem{Bamber:1999zt}
  C.~Bamber {\it et al.},
  %``Studies of nonlinear QED in collisions of 46.6-GeV electrons with  intense
  %laser pulses,''
  Phys.\ Rev.\  D {\bf 60} (1999) 092004.
  %%CITATION = PHRVA,D60,092004;%%


\end{thebibliography}

\end{document}